\documentclass[aps,pre,nofootinbib,showpacs,tightenlines,preprint,titlepage,amsmath]{revtex4}

\usepackage{graphicx,color,epsfig,rotate}
\newcommand{\half}{\textstyle{\frac{1}{2}}}

\newcommand{\cPT}{{\cal PT}}
\newcommand{\cD}{{\cal D}}

\begin{document}

\title{Almost zero-dimensional $\cPT$-symmetric quantum field theories}
\author{Carl~M.~Bender}\email{cmb@wustl.edu}

\affiliation{Department of Physics, Washington University, St. Louis, MO 63130,
USA}

\date{\today}

\begin{abstract}
In 1992 Bender, Boettcher, and Lipatov proposed in two papers a new and unusual
nonperturbative calculational tool in quantum field theory. The objective was to
expand the Green's functions of the quantum field theory as Taylor series in
powers of the space-time dimension $D$. In particular, the vacuum energy for a
massless $\phi^{2N}$ ($N=1,\,2,\,3,\,\ldots$) quantum field theory was studied.
The first two Taylor coefficients in this dimensional expansion were calculated
{\it exactly} and a set of graphical rules were devised that could be used to
calculate approximately the higher coefficients in the series. This approach is
mathematically valid and gives accurate results, but it has not been actively
pursued and investigated. Subsequently, in 1998 Bender and Boettcher discovered
that $\cPT$-symmetric quantum-mechanical Hamiltonians of the form $H=p^2+x^2(ix
)^\epsilon$, where $\epsilon\geq0$, have real spectra. These new kinds of
complex non-Dirac-Hermitian Hamiltonians define physically acceptable
quantum-mechanical theories. This result in quantum mechanics suggests that the
corresponding non-Dirac-Hermitian $D$-dimensional $\phi^2(i\phi)^\epsilon$
quantum field theories might also have real spectra. To examine this hypothesis,
we return to the technique devised in 1992 and in this paper we calculate the
first two coefficients in the dimensional expansion of the ground-state energy
of this complex non-Dirac-Hermitian quantum field theory. We show that to first
order in this dimensional approximation the ground-state energy is indeed real
for $\epsilon\geq0$.
\end{abstract}

\maketitle

\newpage

\section{Introduction}
\label{s1}

Many nonperturbative techniques have been invented to study the properties of
quantum field theory. In this paper we reconsider a little-known nonperturbative
technique that was invented in 1992 by Bender, Boettcher, and Lipatov
\cite{R1,R2} and we use this technique to examine the ground-state energy of a
massless scalar quantum field theory having a self-interaction of the form
$\phi^2(i\phi)^\epsilon$. While the Hamiltonian for this quantum field theory is
complex and obviously not Dirac Hermitian, the Hamiltonian is $\cPT$-symmetric.
The nonperturbative technique developed in Refs.~\cite{R1} and \cite{R2} 
provides support for the conjecture that the ground-state energy for this
strange looking quantum field theory is real.

The nonperturbative technique used in this paper is called {\it dimensional
expansion} and the general idea is to expand the Green's functions of a quantum
field theory as Taylor series in powers of the space-time dimension $D$. The
leading term in such an expansion is easy to determine because it requires only
that one solve the corresponding zero-dimensional version of the quantum field
theory; clearly, this is a trivial exercise. The surprising discovery that was
reported in Refs.~\cite{R1} and \cite{R2} is that the {\it first-order} term
(the coefficient of $D$) in the dimensional expansion can also be calculated
exactly and in closed form. Furthermore, graphical techniques were devised that
can be used to obtain approximately the higher-order terms in the dimensional
expansion.

\begin{figure}[t!]
\vspace{3.2in}
\includegraphics{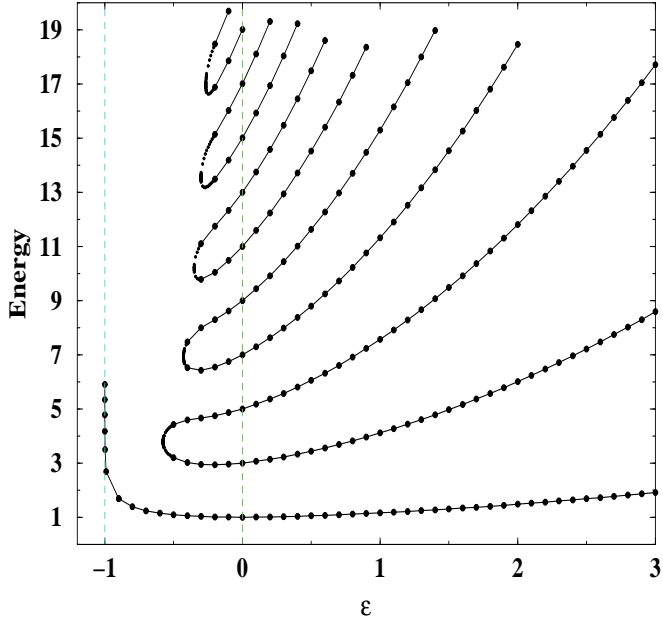}
\caption{Energy levels of the Hamiltonians $H=p^2+x^2(ix)^\epsilon$ in
(\ref{e1}) plotted as functions of the real parameter $\epsilon$. Note that
there are three regions: When $\epsilon\geq0$, the spectrum is discrete, real,
and positive and the energy levels rise with increasing $\epsilon$. The lower
bound of this region, $\epsilon=0$, corresponds to the harmonic oscillator,
whose energy levels are $E_n=2n+1$. When $-1<\epsilon<0$, there are a finite
number of real positive eigenvalues and an infinite number of complex conjugate
pairs of eigenvalues. As $\epsilon$ decreases from $0$ to $-1$, the number of
real eigenvalues decreases; when $\epsilon\leq -0.57793$, the only real
eigenvalue is the ground-state energy. As $\epsilon$ approaches $-1^+$, the
ground-state energy diverges. For $\epsilon\leq -1$ there are no real
eigenvalues. When $\epsilon\geq 0$, the $\cPT$ symmetry is unbroken, but when
$\epsilon<0$ the $\cPT$ symmetry is broken.}
\label{f1}
\end{figure}

In the conventional formulation of quantum mechanics the Hamiltonian $H$ is
assumed to be {\it Dirac Hermitian}, where by this term we mean that $H=H^\dag$
and the symbol $\dag$ represents combined matrix transposition and complex
conjugation. The condition of Dirac Hermiticity allows the matrix elements of
$H$ to be complex, but it guarantees that the eigenvalues of $H$ are real.
However, in 1998 it was shown that it is possible to extend a Hermitian
Hamiltonian into the complex domain while keeping the eigenvalues of the
Hamiltonian real, even though the Hamiltonian is no longer Dirac Hermitian
\cite{R3,R4}. These papers showed how to extend the harmonic oscillator into
the complex domain and thereby obtain the following infinite family of complex
non-Hermitian Hamiltonians having real eigenvalues:
\begin{equation}
H=p^2+x^2(ix)^\epsilon\qquad(\epsilon\geq0).
\label{e1}
\end{equation}
In Refs.~\cite{R3} and \cite{R4} it was argued that the reality of the spectra 
of $H$ in (\ref{e1}) was due to the unbroken $\cPT$ symmetry of these
Hamiltonians for $\epsilon\geq0$. By {\it unbroken} $\cPT$ symmetry, we mean
that the eigenfunctions of the Hamiltonian are also eigenfunctions of the $\cPT$
operator. The eigenvalues of this class of Hamiltonians are plotted in
Fig.~\ref{f1}.

Thus, in quantum mechanics the class of physically allowed Hamiltonians can be
enlarged to include non-Dirac-Hermitian but $\cPT$-symmetric Hamiltonians in
addition to Dirac-Hermitian Hamiltonians, and eigenfunctions (solutions to the
Schr\"odinger equation) are treated as functions of complex coordinates
\cite{R5,R6,R7,R8,R9,R10}. Experimental observations of physical systems
described by complex $\cPT$-symmetric Hamiltonians are now being reported
\cite{R11,R12,R13,R14,R15,R16,R17,R18}.

The fact that $\cPT$-symmetric Hamiltonians can define a conventional theory of
quantum mechanics suggests that it is also possible to extend quantum field
theories into the complex domain \cite{R19}. The quantum-field-theoretic
analogs of the quantum theories described by the Hamiltonians in (\ref{e1}) are
the massless complex scalar quantum field theories whose Euclidean Lagrangians
have the form
\begin{equation}
L=\half(\partial\phi)^2+g\phi^2(i\phi)^\epsilon\qquad(\epsilon\geq0).
\label{e2}
\end{equation}
It has not been proved rigorously that the energies of such quantum field
theories are real but approximate solutions to the Dyson-Schwinger equations
indicate that the poles of the Green's functions are real
\cite{R20,R21,R22,R23,R24}.

This paper is organized very simply. In Sec.~\ref{s2} we review the
dimensional-expansion techniques that were introduced in Refs.~\cite{R1} and
\cite{R2} and show how to apply them to the field theories whose Euclidean
Lagrangians are given in (\ref{e2}). Then, in Sec.~\ref{s3} we make some
concluding remarks and suggest future directions for research.

\section{Dimensional expansions}
\label{s2}

Following Refs.~\cite{R1} and \cite{R2}, we derive a formula for the dimensional
expansion of the {\it logarithmic derivative} $A(\epsilon,D)$ of the free-energy
density $F$ for the Lagrangian $L$ in (\ref{e2}):
\begin{equation}
A(\epsilon,D)\equiv(2+\epsilon)g\frac{dF}{dg}.
\label{e3}
\end{equation}
Note that $A(\epsilon,D)$, as defined in (\ref{e3}), is a dimensionless number
for all $D$ and is independent of the coupling constant $g$; $A(\epsilon,D)$ is
a function only of the dimensionless numbers $\epsilon$ and $D$. We seek a
dimensional expansion of $A(\epsilon,D)$ in the form of a Taylor series in
powers of $D$:
\begin{equation}
A(\epsilon,D)=\sum_{n=0}^\infty a_n(\epsilon)D^n.
\label{e4}
\end{equation}

\subsection{Zeroth-order calculation}
\label{ss2a}

Calculating the first term in the dimensional expansion (\ref{e4}) is easy
because it is obtained from the zero-dimensional version of the field theory:
\begin{eqnarray}
a_0(\epsilon)&=&A(\epsilon,0)\nonumber\\
&=&-(2+\epsilon)g\frac{g}{dg}\log\left\{N\int_C dx\,\exp\left[-gx^2(ix)^\epsilon
\right]\right\}\nonumber\\
&=&(2+\epsilon)g\frac{\int_C dx\,gx^2(ix)^\epsilon\exp\left[-gx^2(ix)^\epsilon
\right]}{\int_C dx\,\exp\left[-gx^2(ix)^\epsilon\right]}\nonumber\\ &=& 1.
\label{e5}
\end{eqnarray}
Thus, the first term in the dimensional expansion (\ref{e4}) is simply unity.

It is important to point out that the contour $C$ for the integrals in
(\ref{e5}) terminates in two Stokes' wedges in the complex-$x$ plane. As
$\epsilon$ increases from 0, these Stokes' wedges (inside of which the integral
converges) rotate downward and become thinner. At $\epsilon=0$ the Stokes'
wedges are centered about the positive-real and negative-real axes and they have
angular opening $90^\circ$. When $\epsilon>0$, the Stokes' wedges are centered
about the two angles
\begin{equation}
\theta_{\rm right}=-\frac{\pi\epsilon}{4+2\epsilon}\quad{\rm and}\quad
\theta_{\rm left}=-\pi+\frac{\pi\epsilon}{4+2\epsilon}.
\label{e6}
\end{equation}
Thus, the two Stokes' wedges are symmetric about the imaginary-$x$ axis. The
opening angles of the Stokes wedges are $\pi/(2+\epsilon)$. Thus, at $\epsilon
=1$ (an $i\phi^3$ theory) the Stokes' wedges are adjacent to and below the real
axes and have angular opening $60^\circ$. When $\epsilon>1$, these wedges lie
below and do not contain the real axis.

This smooth movement of the Stokes' wedges indicates that the field theories
defined by the Lagrangians in (\ref{e2}) can be viewed as being smooth
extensions in the parameter $\epsilon$ of the Dirac-Hermitian
(harmonic-oscillator) Hamiltonian, which lies at $\epsilon=0$, into the complex
non-Dirac-Hermitian domain for which $\epsilon>0$.

\subsection{First-order calculation}
\label{ss2b}

Next, we show how to calculate the coefficient of $D$ in the dimensional
expansion (\ref{e4}). To begin, we add and subtract a mass term in the Euclidean
Lagrangian (\ref{e2}):
\begin{equation}
L=\half(\partial\phi)^2+\half m^2\phi^2+g\phi^2(i\phi)^\epsilon-\half
m^2\phi^2,
\label{e7}
\end{equation}
where $m$ is a mass parameter.

The weak-coupling expansion for the Green's functions of this Lagrangian is in
principle determined by a formal set of Feynman rules:
\begin{eqnarray}
{\rm line:} && \frac{1}{p^2+m^2},\nonumber\\
{\rm 2~vertex:} && m^2,\nonumber\\
(2+\epsilon)~{\rm vertex:} && -i^\epsilon(2+\epsilon)!\,g.
\label{e8}
\end{eqnarray}
Here, the coupling constant $g$ has dimensions of $({\rm mass})^{2+\epsilon-D
\epsilon/2}$. Thus, we choose the mass parameter in (\ref{e7}) equal to the
appropriate power of $g$
\begin{equation}
m=g^{1/(2+\epsilon-D\epsilon/2)}.
\label{e9}
\end{equation} 
It is important to note that if $\epsilon$ is an integer, then there is a
diagrammatic expansion defined by the rules in (\ref{e8}). However, it is not
necessary that $\epsilon$ be an integer; it can be any nonnegative real number.
Thus, unlike the case in Refs.~\cite{R1} and \cite{R2}, the amplitude $-
i^\epsilon(2+\epsilon)!\,g$ does not technically represent a vertex but rather
the continuation of a vertex amplitude off integer values of $\epsilon+2$.

The free-energy density $F$ for the Lagrangian $L$ in (\ref{e7}) is given
by the standard functional-integral representation
\begin{equation}
e^{-FV}=\int_C\cD\phi\,\exp\left(-\int d^Dx\,L\right),
\label{e10}
\end{equation}
where $C$ is the complex contour discussed above, the measure must be properly
normalized, and $V$ represents the volume of space-time in $D$-dimensions. We
can express the right side of (\ref{e10}) as the product of two factors:
\begin{equation}
e^{-FV}=\left[\int_C\cD\phi\,\exp\left(-\int d^Dx\,L_0\right)\right]\times
\left[\frac{\int_C\cD\phi\,\exp\left(-\int d^Dx\,L\right)}
{\int_C \cD\phi\,\exp\left(-\int d^Dx\,L_0\right)}\right].
\label{e11}
\end{equation}
Note that $L_0$ is the free (noninteracting) part of $L$ in (\ref{e7}):
\begin{equation}
L_0=\half(\partial\phi)^2+\half m^2\phi^2.
\label{e12}
\end{equation}

Next we take the logarithm of (\ref{e11}) and express the free energy $F$ as
the sum of two terms:
\begin{equation}
F=F_1+F_2,
\label{e13}
\end{equation}
where
\begin{equation}
F_1=-\frac{1}{V}\log\left[\int_C\cD\phi\,\exp\left(-\int d^Dx\,L_0\right)\right]
\label{e14}
\end{equation}
and
\begin{equation}
F_2=-\frac{1}{V}\left[\frac{\int_C\cD\phi\,\exp\left(-\int d^Dx\,L\right)}
{\int_C \cD\phi\,\exp\left(-\int d^Dx\,L_0\right)}\right].
\label{e15}
\end{equation}
The functional-integral expression for $F_1$ is the free-energy density
associated with the noninteracting Lagrangian in (\ref{e12}). The expression
$F_2$ represents the negative sum of all connected vacuum graphs except for
polygonal loop graphs contributing to the free-energy density of the field
theory described by the Lagrangian $L_0$. This is because the integral in the
denominator of the argument of the logarithm in (\ref{e15}) eliminates from
$F_2$ precisely those one-loop diagrams that contribute to $F_1$. We can in
principle calculate these graphs using the Feynman rules in (\ref{e8}). We now
show how to calculate each of these two quantities.

\vspace{0.2in}
\leftline{\bf Calculation of $F_1$:}

For this calculation we must use the 2 vertex in the Feynman rules in
(\ref{e8}). It is possible to calculate the free-energy density $F_1$
exactly for all dimensions $D$.

The free-energy density $F_1$ is the negative of the total contribution of
all connected vacuum graphs. Since these graphs are constructed from a two-point
vertex, the graphs are all polygons. An $n$-vertex polygon has symmetry
number $1/(2n)$, vertex factor $(-m^2)^n$, and momentum integral
\begin{equation}
(2\pi)^{-D}\int \frac{d^Dp}{p^{2n}}.
\label{e16}
\end{equation}
Hence, apart from an additive divergent zero-point energy contribution that
is independent of $m$, we have
\begin{equation}
F_1=-\sum_{n=1}^\infty\frac{1}{2n}(-m^2)^n(2\pi)^{-D}\int\frac{d^Dp}{p^{2n}}.
\label{e17}
\end{equation}

Each of the integrals in the summation (\ref{e17}) is infrared divergent, but if
we interchange the orders of summation and integration, we see that this
divergence evaporates and we obtain a finite integral representation for $F_1$:
\begin{equation}
F_1=\half(2\pi)^{-D}\int d^Dp\,\log\left(1+\frac{m^2}{p^2}\right).
\label{e18}
\end{equation}
This integral converges for $0<D<2$, but the integral is ultraviolet
divergent at $D=2$.

We can evaluate the integral in (\ref{e18}) straightforwardly by introducing
polar coordinates. The result is 
\begin{equation}
F_1=\frac{1}{D}\left(\frac{m^2}{\pi}\right)^{D/2}\Gamma\left(1-\half D\right).
\label{e19}
\end{equation}
Thus, taking a logarithmic derivative, we find that the contribution of $F_1$ to
$A_1(\epsilon,D)$ is {\it exactly} given by
\begin{equation}
(2\pi)^{-D/2}\Gamma\left(1-\half D\right).
\label{e20}
\end{equation}
The Taylor expansion in powers of $D$ of this quantity begins as follows:
\begin{equation}
1-\half[\log(2\pi)-\gamma]D+{\rm O}(D^2),
\label{e21}
\end{equation}
where $\gamma$ is Euler's constant. These are the first two terms in the
dimensional expansion for the logarithmic derivative of the free energy of the
quantum field theory in (\ref{e12}). The radius of convergence of this
dimensional expansion is 2.

\vspace{0.2in}
\leftline{\bf Calculation of $F_2$:} 

It was shown in Refs.~\cite{R1} and \cite{R2} that while one might imagine that
calculating $F_2$ is difficult because the computation requires the evaluation
of the sum of all vacuum graphs, the calculation to order $D$ is very easy. The 
reason for this is that we only need to calculate the sum of the vacuum graphs
for the case of zero dimensions if we want to know $A(\epsilon,D)$ to order $D$.
This is true because the contribution of each vacuum graph is a finite quantity
having dimensions of $m^D$ (the dimensions of an energy density) and each graph
is finite in the limit as $D\to0$, unlike the expression for $F_1$ in
(\ref{e19}). Thus, each vacuum graph has the general form 
\begin{equation}
m^D[{\rm constant}+{\rm O}(D)].
\label{e22}
\end{equation}
Consequently, when we calculate the logarithmic derivative of each graph in
$F_2$ to obtain $A(\epsilon,D)$, the factor of $m^D$ contributes a factor of
$D$. The vacuum graphs need only be calculated for the special case $D=0$.

To verify the form in (\ref{e22}), we argue as follows: For every Feynman graph
we must evaluate a $D$-dimensional momentum integral as required by the Feynman
rules in (\ref{e8}). The integrand $f(p)$ of this integral is a combination of
propagators of the form $1/(p^2+m^2)$. In the limit as $D\to0$, we may replace 
$f(p)$ by $f(0)$ because there is no infrared divergence in any graph. In fact,
the value of the integral as $D\to0$ is just $f(0)$.

It follows that we can express $F_2$ for small $D$ as a ratio of dimensionless
integrals multiplied by the mass factor $m^D$:
\begin{equation}
F_2=-m^D\log\frac{\int_C dx\,\exp\left[-x^2(ix)^\epsilon\right]}
{\int_{-\infty}^{\infty}dx\,\exp\left(-\half x^2\right)}.
\label{e23}
\end{equation}
We evaluate the integral in the numerator of (\ref{e23}) by reducing it to a
real integral along the centers of the Stokes' wedges. The result is that
\begin{equation}
F_2=-m^D\log\left[(2/\pi)^{1/2}\Gamma\left(1+\frac{1}{2+\epsilon}\right)\cos
\left(\frac{\pi\epsilon}{4+2\epsilon}\right)\right].
\label{e24}
\end{equation}
This gives the order ${\rm O}(D)$ contribution from $F_2$ to $A(\epsilon,D)$.

\vspace{0.2in}
\leftline{\bf Combination of $F_1$ and $F_2$:} 

Finally, we combine the contributions to $A(\epsilon,D)$ from $F_1$ in
(\ref{e21}) and from $F_2$ in (\ref{e24}). The result is that the
dimensional expansion has the form
\begin{equation}
A(\epsilon,D)=1-\half
D\left[4e^{-\gamma}\Gamma^2\left(1+\frac{1}{2+\epsilon}\right)\cos^2\left(
\frac{\pi\epsilon}{4+2\epsilon}\right)\right]D+{\rm O}\left(D^2\right),
\label{e25}
\end{equation}
and we emphasize that the coefficients of $D^0$ and $D^1$ in this expansion are
exact.

\section{Final remarks and future research}
\label{s3}

In this paper we have used the dimensional-expansion techniques proposed by
Bender, Boettcher, and Lipatov in 1992 to find exactly the first two terms of
the dimensional expansion (\ref{e25}) for the ground-state energy density of the
Euclidean quantum field theory defined by the complex, non-Dirac-Hermitian
Lagrangian in (\ref{e2}). The reality of the coefficients in this expansion
suggests that the entire expansion may well have real coefficients. 

We do not know how to calculate the higher coefficients in the dimensional
expansion exactly. However, graphical techniques were devised in Refs.~\cite{R1}
and \cite{R2} that can be used to find accurate approximations to the higher
coefficients. This would be a worthy objective for a research project, and an
important step in showing that non-Dirac-Hermitian Lagrangians can define
physically acceptable quantum field theories.

\begin{acknowledgments}
I thank the U.S.~Department of Energy for financial support.
\end{acknowledgments}

\end{document}